\documentclass[12pt]{iopart}
\input{amssym}
\usepackage{epsfig}
\global\arraycolsep=2pt

\newcommand{\ben}{\begin{equation*}}
\newcommand{\een}{\end{equation*}}
\newcommand{\bean}{\begin{eqnarray*}}
\newcommand{\eean}{\end{eqnarray*}}
\newcommand{\bnabla}{\mbox{\boldmath{$\nabla$}}}
\newcommand{\nn}{\nonumber}
\newcommand{\be}{\begin{equation}}
\newcommand{\ee}{\end{equation}}
\newcommand{\bea}{\begin{eqnarray}}
\newcommand{\eea}{\end{eqnarray}}

\begin{document}

\title{Gravitational and Inertial Mass of Casimir Energy}
\author{Kimball A Milton$^1$, Stephen A Fulling$^2$,
Prachi Parashar$^1$, August Romeo$^3$, K V Shajesh$^1$,  Jeffrey A Wagner$^1$}

\address{$^1$Oklahoma Center for High Energy Physics and
H. L. Dodge Department of Physics and Astronomy, University of Oklahoma,
Norman, OK 73019-2061 USA\\
$^2$Departments of Mathematics and Physics, Texas A\&M University,
College Station, TX 77843-3368 USA\\ 
$^3$Societat Catalana de F\'\i sica, Laboratori de F\'\i sica Matem\`atica
(SCF-IEC), 08028 Barcelona, Catalonia, Spain}
\ead{milton@nhn.ou.edu}

\begin{abstract}
It has been demonstrated, using variational methods, 
that quantum vacuum energy gravitates according
to the equivalence principle, at least for the finite Casimir energies
associated with perfectly conducting parallel plates.
This conclusion holds independently of the orientation of the plates. 
 We review these arguments and add further
support to this conclusion by considering parallel semitransparent plates,
that is, $\delta$-function potentials, acting on a massless scalar field,
in a spacetime defined by Rindler coordinates.
We calculate the force on systems consisting of one or
two such plates undergoing acceleration perpendicular to the plates.
In the limit of small
acceleration we recover (via the equivalence principle)
the situation of weak gravity, and find that the gravitational force on
the system is just $M\mathbf{g}$, where $\mathbf{g}$
is the gravitational acceleration
and $M$ is the total mass of the system, consisting of the mass of the plates
renormalized by the Casimir energy of each plate separately, plus the energy
of the Casimir interaction between the plates.  This reproduces the previous
result in the limit as the coupling to the $\delta$-function potential
approaches infinity.  Extension of this latter work to arbitrary orientation of
the plates, and to general compact quantum vacuum energy configurations,
is under development.
\end{abstract}


\section{Introduction}
The subject of Quantum Vacuum Energy (the Casimir effect) dates from the
same year as the discovery of renormalized quantum electrodynamics, 1948,
and suggests that the assertion that zero-point energy is not
observable is invalid. (For a contrary viewpoint see Ref.~\cite{Jaffe:2005vp}.)
On the other hand, because of the severe divergence structure
of the theory, controversy has surrounded it from the beginning.
Sharp boundaries give rise to divergences in the local energy density near
the surface, which may make it impossible to extract meaningful self-energies
of single objects, such as the perfectly conducting sphere considered by
Boyer~\cite{Boyer:1968uf}.  These objections have recently
 been most forcefully presented by Graham,
et al.~\cite{Graham:2003ib} and Barton~\cite{barton}, but they date 
back to Deutsch and Candelas \cite{Deutsch:1978sc,Candelas:1981qw}. 
In fact,
it now appears that these surface divergences can be dealt with successfully
in a process of renormalization, and that finite self-energies in the sense
of Boyer, may be extracted  \cite{CaveroPelaez:2006kq,CaveroPelaez:2006rt}.

But the most troubling aspect of local energy divergences is in the coupling
to gravity.  The source of gravity is 
the local energy-momentum tensor, and such
surface divergences promise serious difficulties.  
As a prolegomenon to studying
such questions, we here address in \sref{sec1} a simpler question: How does 
the completely finite Casimir energy of a pair of parallel conducting plates 
respond to gravity? (We'll address divergences in \sref{sec2}.)
The question, and its answer, turn out to be surprisingly less straightforward
than the reader might suspect!  (For a complementary view on the gravitational
effects of Casimir energy, see the contribution to these Proceedings by
S A Fulling et al.)

\section{Variational method}
\label{sec1}
\subsection{Casimir stress tensor for parallel plates}
Brown and Maclay \cite{brown} showed that, for parallel perfectly
conducting plates separated by a distance $a$ in the $z$-direction, 
the electromagnetic stress
tensor acquires the vacuum expectation value between the plates
\be
\langle T^{\mu\nu}\rangle
=\frac{\mathcal{E}_c}a \mbox{diag} (1, -1, -1, 3),
\quad\mathcal{E}_c=-\frac{\pi^2}{720 a^3}\hbar c.\ee
Outside the plates the value of $\langle T^{\mu\nu}\rangle=0$.  
Because
there are some subtleties here, let us review the argument for the
case of a conformally coupled scalar (the electromagnetic case differs
by a factor of two).  Actually, the result between the plates,
$0<z<a$ is given
in great detail in Ref.~\cite{casbook} ($\gamma$ is the conformal parameter):
\be
\langle T^{\mu\nu}\rangle=
(u_0+u)\mbox{diag}(1,-1,-1,3)
+(1-6\gamma)g(z)\mbox{diag}(1,-1,-1,0),\label{bmt}
\ee
where
\be
u_0=-\frac1{12\pi^2}\int_0^\infty \rmd\kappa\,\kappa^3,\quad
u=-\frac{\pi^2}{1440 a^4}.
\ee
Note that $u_0$ is a divergent constant, independent of $a$,
and is present (as we shall see) both inside and outside the plates,
so it does not contribute to any observable force or energy (the force
on the plates is given by the discontinuity of $\langle T_{zz}\rangle$),
and so may be simply disregarded (as long as we are not concerned with
dark energy).  But see below!
Similarly, the term involving the Hurwitz zeta function,
\be
g(z)=-\frac1{16\pi^2a^4}[\zeta(4,z/a)+\zeta(4,1-z/a)],
\ee
which exhibits the universal surface
divergence near the plates, 
\be
g(z)\sim -\frac1{16\pi^2 z^4},\quad z\to 0+,
\ee
is also unobservable (if we disregard gravity)
because it does not
contribute to the force on the plates, nor does
it contribute to the total energy, since the integral over $g(z)$ between
the plates is independent of the plate separation.  Of course, the best
way to eliminate that term is to choose the conformal value $\gamma=1/6$.

Since the exterior calculation does not appear to be referred to in 
Ref.~\cite{casbook}, let us sketch the calculation here:  Consider parallel 
Dirichlet plates at $z=0$ and $z=a$.
The reduced Green's function satisfies
\be
\left(-\frac{\rmd^2}{\rmd z^2}+\kappa^2\right)g(z,z')=\delta(z-z'),
\ee
where $\kappa^2=k^2-\omega^2=k^2+\zeta^2$.  The solution for $z, z'<0$ is
\be
g(z,z')=-\frac1\kappa \rme^{\kappa z_<}\sinh \kappa z_>.
\ee
It is very straightforward to calculate the one-loop expectation value of
the stress tensor from
\be
\rmi\langle T^{\mu\nu}\rangle=\left(\partial^\mu\partial^{\prime\nu}-\frac12
g^{\mu\nu}\partial^\lambda\partial_\lambda'\right)G(x,x')\bigg|_{x'=x}
-\gamma(\partial^\mu\partial^\nu-g^{\mu\nu}\partial^2)G(x,x).
\ee
After integrating over $\omega=\rmi\zeta$ and $\mathbf{k}$,
we find the result ($z<0$)
\be
\langle T^{\mu\nu}\rangle=u_0\mbox{diag}(1,-1,-1,3)-\frac{(1-6\gamma)}{16\pi^2
|z|^4}\mbox{diag}(1,-1,-1,0).
\ee
This is exactly as expected.  The $u_0$ term is the same as inside the
box, so is just the vacuum value, and the second term is the universal surface
divergence (independent of plate separation), 
which can be eliminated by setting $\gamma=1/6$. 

 Thus, we conclude that the
physical stress tensor VEV is just that found by Brown and Maclay:
\be
\langle T^{\mu\nu}\rangle =u\,\mbox{diag}(1,-1,-1,3)\theta(z)\theta(a-z).
\label{gravfree}
\ee
in terms of the usual step function.

\subsection{Variational principle}
Now we address the question of the gravitational interaction of
this Casimir apparatus~\cite{Fulling:2007xa}.
It seems this question can be most simply answered through
use of the gravitational definition of the energy-momentum tensor,
\be
\delta W_m\equiv -\frac12\int(\rmd x) \sqrt{-g}\,\delta g^{\mu\nu}T_{\mu\nu}
=\frac12\int(\rmd x) \sqrt{-g}\,\delta g_{\mu\nu}T^{\mu\nu}.\label{var}
\ee
For a weak field, $g_{\mu\nu}=\eta_{\mu\nu}+2h_{\mu\nu}$
(Schwinger's definition \cite{js} of $h_{\mu\nu}$).
So if we think of turning
on the gravitational field as the perturbation, we can ignore $\sqrt{-g}$.
The gravitational energy, for a static situation, is therefore given by
($\delta W=-\int dt\,\delta E$)
\be
E_g=-\int (\rmd\mathbf{x}) h_{\mu\nu}T^{\mu\nu}.\label{ge}
\ee
We can use the gravity-free electromagnetic Casimir stress tensor
\eref{gravfree}, 
with $u$ now replaced by $\mathcal{E}_c/a$ for the electromagnetic situation.

\begin{figure}
\centering
\epsfig{figure=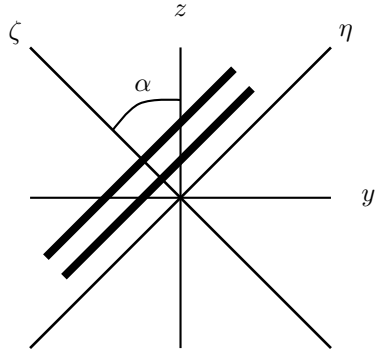}
\caption{Relation between two Cartesian coordinate frames: One attached to
the earth $(x,y,z)$, where $-z$ is the direction of gravity, and one attached
to the parallel-plate Casimir apparatus $(\zeta,\eta,\chi)$, where $\zeta$ is
in the direction normal to the plates.  The parallel plates are
indicated by the heavy lines parallel to the $\eta$ axis.  The $x=\chi$
axis is perpendicular to the page.}
\label{fig1}
\end{figure}

We now use the metric \cite{calloni,bimonte}
\be
g_{00}=-\left(1+2gz\right),\quad g_{ij}=\delta_{ij}.\label{const}
\ee
This is appropriate for a constant gravitational
field. (But see below.) Let us consider a Casimir apparatus of parallel plates
separated by a distance $a$, with transverse dimensions $L\gg a$.
Let the apparatus be oriented at an angle $\alpha$ with respect to the
direction of gravity, as shown in \fref{fig1}.
Let us take the Cartesian coordinate
system attached to the earth to be $(x, y, z)$, where, as noted above,
$z$ is the direction of $-\mathbf{g}$.  Let the Cartesian coordinates
associated with the Casimir apparatus be $(\zeta,\eta,\chi)$, where $\zeta$
is normal to the plates, and $\eta$ and $\chi$ are parallel to the plates.
The relation between the two sets of coordinates is 
\be
 z=\zeta\cos\alpha+\eta\sin\alpha,\quad
y=\eta\cos\alpha-\zeta\sin\alpha,\quad
x=\chi.
\ee
Let the center of the apparatus be located at $(\zeta_0,\eta=0,\chi=0)$.

Now we calculate the gravitational energy
\bea
E_g&=&\int (\rmd\mathbf{x})gzT^{00}
=\frac{\mathcal{E}_c}a gL\int_{-L/2}^{L/2}\rmd\eta
\int_{\zeta_0-a/2}^{\zeta_0+a/2}\rmd\zeta(\zeta\cos\alpha+\eta\sin\alpha)\nn\\
&=&\frac{g\mathcal{E}_c}{a}L^2\cos\alpha a\zeta_0+K,
\eea
where $K$ is a constant, independent of $\zeta_0$.
Thus, the gravitational force per area on the apparatus is independent of
orientation
\be
\frac{F}A=-\frac{\partial E_g}{A\partial z_0}=-\frac\epsilon{2a}\mathcal{E}_c
=-g\mathcal{E}_c,
\quad z_0=\zeta_0\cos\alpha,
\label{f0}
\ee
a small upward push. Here $\epsilon=2ga$ is a measure of the gravitational
force relative to the Casimir force.
 Note that on the earth's surface, 
the dimensionless number $\epsilon$ is very small.  For a plate separation
of $1 \mu$m,
\be
\epsilon=\frac{2ga}{c^2}=2.2\times 10^{-22},
\ee
so the considerations here would appear to be only of theoretical interest.
The effect is far smaller than the Casimir forces between the plates.

It is somewhat simpler to use the energy formula to calculate the 
force by considering the variation in the gravitational energy
directly, as we can illustrate by considering a mass point at the origin:
\be
T^{\mu\nu}=m\delta(\mathbf{r})\delta^{\mu0}\delta^{\nu0}.\label{masspt}
\ee
If we displace the particle rigidly upward by an amount $\delta z_0$, the
change in the metric is $\delta h_{00}=-g \delta z_0$.
This implies a change in the energy, exactly as expected:
\be
\delta E_g=-mc^2\left(-g \delta z_0\right)=mg\delta z_0.
\ee

Now we repeat this calculation for the Casimir apparatus. The gravitational
force per area on the rigid apparatus is 
$\frac{F}A=-\frac{\delta \mathcal{E}_g}{\delta z_0}=- g\mathcal{E}_c$,
 again the same result found in \eref{f0}, which agrees with the second
result found by Calloni et al.~\cite{calloni} but is
1/4 that found by Bimonte et al.~\cite{bimonte},
who reproduce the first result of Ref.~\cite{calloni}. 
Our result is consistent with the
principle of equivalence, and with one result of
Jaekel and Reynaud \cite{jaekel}.

\subsection{Alternative calculation}

As in electrodynamics, we should be able to proceed, starting
from the definition of the field
\be
\delta W=\int (\rmd x) \delta T^{\mu\nu} h_{\mu\nu}.\label{2ndmethod}
\ee
Again, check this for the force on a mass point, with stress tensor
given by \eref{masspt},
so
\be
\delta E_g=- \int
(\rmd\mathbf{r}) m \left(-\delta\mathbf{r}\cdot\bnabla\right)
\delta(\mathbf{r})h^{00}=- m\delta\mathbf{r}\cdot \bnabla h^{00}.
\ee
Since $h^{00}=-gz$, we conclude
$-\frac{\delta E}{\delta \mathbf{r}}=\mathbf{F}=-mg \hat z$.

For the constant field the force on a Casimir apparatus
is obtained from the change in the energy density
\be
T^{00}=\frac{\mathcal{E}_c}a\theta(a/2-\zeta+\zeta_0)\theta(\zeta-\zeta_0+a/2),
\ee
that is, recalling that $z_0=\zeta_0\cos\alpha$,
\be
\delta T^{00}=\frac{\mathcal{E}_c}a\delta z_0\frac1{\cos\alpha}
\left[\delta (\zeta-\zeta_0-a/2)-\delta (\zeta-\zeta_0+a/2)\right],
\label{varst}
\ee
which yields a result identical to \eref{f0}
[$h^{00}=-g (\zeta\cos\alpha+\eta\sin\alpha)$]
\be
-\frac{\delta E_g}{A\delta z_0}=\frac{F}A=\frac{\mathcal{E}_c}a
\frac1{\cos\alpha}
 h^{00}\bigg|_{\zeta =\zeta_0-a/2}^{\zeta=\zeta_0+a/2}=
-g\mathcal{E}_c.
\ee

\subsection{Metric near the surface of the earth}
However, the above metric \eref{const}, 
while sufficing for massive Newtonian objects,
might seem inappropriate for photons.  Rather, shouldn't we
 use the perturbation of
the Schwarzschild metric, which for weak fields ($GM/r\ll1$)
 is in isotropic
coordinates \cite{weinberg}:
\be
\rmd s^2=-\left(1-\frac{2GM}r\right)c^2\rmd t^2+\left(1+\frac{2GM}r\right)
\rmd\mathbf{r}^2?
\ee
If we expand this
 a short distance $z$ above the earth's surface, of radius $R$, we find
\be
g_{00}\approx -\left(1-\frac{2GM}R + 2gz\right),\quad
 g_{ij}\approx\delta_{ij}\left(1+\frac{2GM}R-2gz\right).\label{ic}
\ee

Now, for our Casimir apparatus shown in \fref{fig1},
each component of the Casimir stress tensor contributes with equal weight:
\be
-\frac{\delta E_g}{A\delta z_0}=-ga
\left(T^{00}+T^{11}+T^{22}+T^{33}\right)
=-2g\mathcal{E}_c,\label{r1}
\ee
since $T=T^\lambda{}_\lambda=0$,
which is twice the previous result.  
Note that again the result is independent of $\alpha$.
If instead, we use the second method we have 
\be
\fl\delta T^{\mu\nu}=-\delta z\frac{\mathcal{E}_c}{a}(1,-1,-1,3)
\frac1{\cos\alpha}
\left[\delta\left(\zeta-\zeta_0+\frac{a}2\right)-
\delta\left(\zeta-\zeta_0-\frac{a}2\right)\right],
\ee
so, again we get the same result:
\be
\fl-\frac{\delta E_g}{\delta z_0}=-
\frac{\mathcal{E}_c}a\int\frac{(\rmd\mathbf{r})}{\cos\alpha}
\left[\delta\left(\zeta-\zeta_0+\frac{a}2\right)-\delta
 \left(\zeta-\zeta_0-\frac{a}2\right)\right]
\left(-2g z\right)=F=-A\,2g
\mathcal{E}_c,
\ee
where $z=\zeta\cos\alpha+\eta\sin\alpha$.

We might think we would 
be able to obtain the same result using the original Schwarzschild
coordinates, where
$
h_{00}=-gz,\quad h_{\rho\rho}=-gz,
$
and all other components of $h_{\mu\nu}$ are zero.  However, now if we use
the first method above, the result is proportional to 
$T^{00}+T^{\rho\rho}=\frac{\mathcal{E}_c}a 4\cos^2\alpha,$
which implies (fortuitously) Bimonte et al.'s earlier result 
\cite{bimonte} for $\alpha=0$:
$\frac{F}A=-4g\mathcal{E}_c\cos^2\alpha$.

\subsection{Gauge noninvariance}
The reason we get different answers in different coordinate systems
is that our starting point is not gauge invariant.
Under a coordinate redefinition, which for weak fields is a gauge 
transformation of $h_{\mu\nu}$ \cite{js},
$h_{\mu\nu}\to h_{\mu\nu}+\partial_\mu\xi_\nu+\partial_\nu\xi_\mu$,
where $\xi_\mu$ is a vector field,  the interaction $W$ 
is invariant only if the
stress tensor is conserved, $\partial_\mu T^{\mu\nu}=0.$  Otherwise, there
is a change in the action,
$\Delta W=-2\int (\rmd x)\xi_\nu \partial_\mu T^{\mu\nu}.$

Now in our case (where we make the finite size of the plate explicit,
but ignore edge effects because $L\gg a$)
\bea
T^{\mu\nu}&=&\frac{\mathcal{E}_c}a\mbox{diag}(1,-1,-1,3)\theta
\left(\zeta-\zeta_0+\frac{a}2\right)
\theta\left(\frac{a}2-\zeta+\zeta_0\right)\nn\\
&&\quad\times\theta(\eta+L/2)\theta(L/2-\eta)\theta(\chi+L/2)\theta(L/2-\chi).
\label{explicitt}
\eea
Thus the nonzero components of $\partial_\mu T^{\mu\nu}$ are
\numparts
\bea
\partial_\mu T^{\mu\zeta}&=&\frac{3\mathcal{E}_c}a
\left[\delta(\zeta-\zeta_0+a/2)-\delta(\zeta-\zeta_0-a/2)\right]\theta\dots,\\
\delta_\mu T^{\mu\eta}&=&-\frac{\mathcal{E}_c}a\left[\delta(\eta+L/2)-\delta
(\eta-L/2)\right]\theta\dots,\\
\delta_\mu T^{\mu\chi}&=&-\frac{\mathcal{E}_c}a\left[\delta(\chi+L/2)-\delta
(\chi-L/2)\right]\theta\dots,
\eea
\endnumparts
where $\theta\dots$ refer to the remaining step functions.
Therefore, the change in the energy obtained from $\Delta W$ is
\bea
\Delta E_g&=&\frac{6\mathcal{E}_c}{a}\int\rmd\eta\, \rmd\chi \left[\xi_\zeta(
\zeta_0-a/2,\eta,\chi)-\xi_\zeta(\zeta_0+a/2,\eta,\chi)\right]\nn\\
&&\quad\mbox{}-\frac{2\mathcal{E}_c}a\int\rmd\zeta \,\rmd\chi 
\left[\xi_\eta(\zeta,-L/2,\chi)-\xi_\eta(\zeta,L/2,\chi)\right]\nn\\
&&\quad\mbox{}-\frac{2\mathcal{E}_c}a\int \rmd\zeta\, \rmd\eta 
\left[\xi_\chi(\zeta,
\eta, -L/2)-\xi_\chi(\zeta,\eta,L/2)\right].\label{de}
\eea

\subsection{Fermi coordinates}
Since we have demonstrated that the gravitational force on a Casimir
apparatus is not a gauge-invariant concept, we must ask if there is
any way to extract a physically meaningful result. There seem to be
two possible ways to proceed.  Either we add another interaction, say
a fluid exerting a pressure on the plates, resulting in a total stress
tensor that is conserved, or we find a physical basis for believing that
one coordinate system is more realistic than another.
The former procedure
is undoubtedly more physical, but will yield model dependent results.
The latter apparently has a natural solution.

A Fermi coordinate system is the general relativistic generalization of
an inertial coordinate frame.  Such a system has been given by 
Marzlin \cite{marzlin} for a resting observer in the field of a static mass
distribution.  
It is actually {\it a priori} obvious that in such a system
$g_{ij}$ is quadratic in the distance from the observer.
Thus the ``constant field metric'' is simply
the Fermi coordinate metric for a gravitating body,
\be
\rmd s^2=-(1+2gz)\rmd t^2+\rmd\mathbf{r'}^2.\label{fc}
\ee
Thus, coordinate lengths don't depend on $z$.  The metric \eref{const}
is indeed appropriate, and the corresponding
gravitational force is therefore given by the result found in that case,
$F/A=-g\mathcal{E}_c$, as in \eref{f0}.

\subsection{Gauge transformation}
Now we can use the method described in \eref{de} to transform
the energy in isotropic coordinates to that in Fermi coordinates.  We compute 
the additional gravitational energy, in terms
of the gauge field $\xi_\mu$, which carries us from isotropic coordinates
to Fermi coordinates,
$h_{\mu\nu}^F=h_{\mu\nu}^I+\partial_\mu\xi_\nu+\partial_\nu\xi_\mu$.
Here from \eref{ic} and \eref{fc}
\be
h_{00}^I=-gz,\quad h^I_{ij}=-gz\delta_{ij},
\quad h_{00}^F=-gz,\quad h^F_{ij}=0.
\ee
The gauge field turns out to be
\numparts
\bea
\xi_\zeta&=&\frac12g \left(\frac12\zeta^2\cos\alpha+\zeta\eta\sin\alpha
\right)+f(\eta,\chi),\\
\xi_\eta&=&\frac12g \left(\zeta\eta\cos\alpha+\frac12\eta^2\sin\alpha
\right)+g(\zeta,\chi),\\
\xi_\chi&=&\frac12g \left(\zeta\cos\alpha+\eta\sin\alpha\right)\chi
+h(\zeta,\eta),
\eea
\endnumparts
where the functions $f$, $g$, and $h$ are irrelevant.
Substituting this into the expression for $\Delta E_g$, \eref{de}, we obtain
\bea
\fl\Delta E_g&=&\frac{6 \mathcal{E}_c}a
\int_{-L/2}^{L/2}\rmd\eta\int_{-L/2}^{L/2}\rmd\chi
\frac14 g\cos\alpha\left(-2\zeta_0a\right)
-\frac{2\mathcal{E}_c}a\int_{\zeta_0-a/2}^{\zeta_0+a/2}\rmd\zeta
\int_{-L/2}^{L/2}\rmd\chi\frac12g\cos\alpha(-L)\zeta\nn\\
\fl
&&\quad\mbox{}-\frac{2\mathcal{E}_c}a\int_{\zeta_0-a/2}^{\zeta_0+a/2}\rmd\zeta
\int_{-L/2}^{L/2}\rmd\eta\frac12g(\zeta\cos\alpha+\eta\sin\alpha)(-L)\nn\\
\fl&=&-Ag\mathcal{E}_c\zeta_0\cos\alpha=-Ag\mathcal{E}_c z_0,
\eea
which when differentiated with respect to $z_0$ gives an additional force,
$-\frac{\delta\Delta E_g}{A\delta z_0}=\frac{\Delta F}A=g\mathcal{E}_c$.
When this is added to the isotropic force \eref{r1}, we obtain the Fermi force,
\be
\frac{F^I+\Delta F}A=-2g\mathcal{E}_c+g\mathcal{E}_c=-g\mathcal{E}_c=
\frac{F^F}A,
\ee
as given in \eref{f0}.  This answer is the second one given in 
Calloni et al.~\cite{calloni},
but is not referred to in the 2006 Bimonte et al.\ paper \cite{bimonte}.
Those authors have now modified their analysis to agree with ours
\cite{Bimonte:2007zt}.

\section{Rindler coordinates}
\label{sec2}
We now turn to the consideration of the Casimir apparatus undergoing
uniform acceleration \cite{Milton:2007ar}.
Relativistically, uniform acceleration is described by hyperbolic motion
\be
t=\xi\sinh\tau,\quad z=\xi\cosh\tau,
\ee
where $\xi^{-1}$ is the proper acceleration, which corresponds to the metric
\be
\rmd t^2-\rmd z^2=\xi^2\rmd \tau^2-\rmd\xi^2.
\ee
The d'Alembertian operator takes on cylindrical form
\be
-\left(\frac\partial{\partial t}\right)^2
+\left(\frac\partial{\partial z}\right)^2
=-\frac1{\xi^2}\left(\frac\partial{\partial\tau}\right)^2+\frac1\xi
\frac\partial{\partial\xi}\left(\xi\frac\partial{\partial\xi}\right).
\ee

\subsection{Single accelerated plate}
For a single semitransparent
plate at $\xi_1$, the Green's function can be written as
\be
G(x,x')=\int\frac{\rmd\omega}{2\pi}\frac{\rmd^2 k_\perp}{(2\pi)^2}
\rme^{-i\omega(\tau-\tau')}
\rme^{i\mathbf{k_\perp\cdot(r-r')_\perp}}g(\xi,\xi'),
\ee
where the reduced Green's function satisfies ($k=|\mathbf{k_\perp}|$)
\be
\left[-\frac{\omega^2}{\xi^2}+\frac1\xi\frac\partial{\partial\xi}\left(\xi\frac
\partial{\partial\xi}\right)+k^2+\mu\delta(\xi-\xi_1)\right]g=
\frac1\xi\delta(\xi-\xi'),
\ee
which we recognize as just the 
semitransparent cylinder problem with $m\to \zeta=-\rmi\omega$ and
$\kappa\to k$. Thus, the Green's function for a single plate is
\numparts
\bea
g(\xi,\xi')&=&I_\zeta(k\xi_<)K_\zeta(k\xi_>)-\frac{\mu \xi_1 K_\zeta^2(k\xi_1)
I_\zeta(k\xi)I_\zeta(k\xi')}{1+\mu \xi_1 I_\zeta(k\xi_1)K_\zeta(k\xi_1)},
\quad\xi,\xi'<\xi_1,\\
&=&I_\zeta(k\xi_<)K_\zeta(k\xi_>)-\frac{\mu \xi_1 I_\zeta^2(k\xi_1)
K_\zeta(k\xi)K_\zeta(k\xi')}{1+\mu \xi_1 I_\zeta(k\xi_1)K_\zeta(k\xi_1)},
\quad\xi,\xi'>\xi_1.
\eea
\endnumparts
where the strong coupling limit,
$\mu\to\infty$, corresponds to Dirichlet boundary conditions.

\subsection{Minkowski-space limit}
If we use the uniform asymptotic expansion (UAE), based on the
limit
\be
\xi\to\infty,\quad \xi_1\to\infty, \quad \xi-\xi_1 \mbox{ finite },\quad
\zeta=\hat\zeta\xi_1\to\infty,\quad \hat\zeta \mbox{ finite },
\ee
we recover the Green's function for a single plate in Minkowski space,
\be
\xi_1 g(\xi,\xi')\to  \frac{\rme^{-\kappa|\xi-\xi'|}}{2\kappa}
-\frac\mu{2\kappa}\frac{\rme^{-\kappa(|\xi-\xi_1|+
|\xi'-\xi_1|)}}{\mu+2\kappa},
\ee
where $\kappa=\sqrt{k^2+\zeta^2}$, $\omega=\rmi\zeta$. 

\subsection{Energy-momentum tensor}
The canonical energy-momentum for a scalar field is given by
$T_{\mu\nu}=\partial_\mu\phi\partial_\nu\phi
+g_{\mu\nu}\frac1{\sqrt{-g}}\mathcal{L}$,
where the Lagrange density includes the $\delta$-function potential.  Using
the equations of motion the energy density is
\be
\fl T_{00}=\frac12\left(\frac{\partial\phi}{\partial\tau}\right)^2-\frac12\phi
\frac{\partial^2}{\partial\tau^2}\phi+\frac\xi 2\frac\partial{\partial\xi}
\left(\phi\xi\frac\partial{\partial\xi}\phi\right)+\frac{\xi^2}2\bnabla_\perp
\cdot(\phi\bnabla_\perp\phi).\label{enden}\ee
The force density is given by
\be
f_\lambda=-\frac1{\sqrt{-g}}\partial_\nu(\sqrt{-g}T^\nu{}_\lambda)
+\frac12T^{\mu\nu}\partial_\lambda g_{\mu\nu},
\ee
or
\be
f_\xi=-\frac1\xi\partial_\xi(\xi T^{\xi\xi})-\xi T^{00}.
\ee
When we integrate over all space to get the (``coordinate'') force (per area), 
the first term is a surface term which does not contribute:
\be
\mathcal{F}=\int \rmd\xi\, \xi \,f_\xi=-\int\frac{\rmd\xi}{\xi^2}T_{00},
\ee
which when multiplied by the gravitational acceleration $g$ is the
gravitational force/area on the Casimir energy.
Using the expression \eref{enden} for the energy density, and rescaling 
$\zeta=\hat\zeta\xi$,
we see that the gravitational force is merely
\be
\mathcal{F}=\int \rmd\xi \,\xi\int\frac{\rmd\hat\zeta \, \rmd^2k}{(2\pi)^3}
\hat\zeta^2 
g(\xi,\xi).
\ee

This result is an immediate consequence of the general formula
\be
E_c=-\frac1{2\rmi}\int(\rmd\mathbf{r})\int\frac{\rmd\omega}{2\pi}
2\omega^2\mathcal{G}(\mathbf{r,r}),
\ee
in terms of the frequency transform of the Green's function,
\be
G(x,x')=\int_{-\infty}^\infty\frac{\rmd\omega}{2\pi}
\rme^{-\rmi\omega(t-t')}\mathcal{G}(\mathbf{r,r'}).
\ee

\subsection{Force on single plate}
Alternatively, we can start from the following formula for the force
density for a single semitransparent plate,
\be
f_\xi=\frac12\phi^2\partial_\xi \mu\delta(\xi-\xi_1),
\ee
or, in terms of the Green's function,
\be
\mathcal{F}=-\mu\frac12
\int\frac{\rmd\zeta\,\rmd^2k}{(2\pi)^3}\partial_{\xi_1}[\xi_1 g(\xi_1,\xi_1)].
\ee
For example, the force on a single plate is given by
\be
\mathcal{F}=-\partial_{\xi_1} \frac12\int\frac{\rmd\zeta\,\rmd^2k}{(2\pi)^2}\ln
[1+\mu\xi_1 I_\zeta(k\xi_1)K_\zeta(k\xi_1)],
\ee
Expanding this about some arbitrary point $\xi_0$, with $\zeta=\hat\zeta\xi_0$,
and using the UAE, we get ($a$ is an arbitrary scale to make $y$ dimensionless)
\be
\mathcal{F}=-\frac1{96\pi^2 a^3}\int_0^\infty\frac{\rmd y\,y^2}{1+y/\mu a},
\label{oneplate}
\ee 
which is just the negative of the (divergent)
quantum vacuum energy of a single plate.

\subsection{Two accelerated plates}

For two plates at $\xi_1$, $\xi_2$, for $\xi, \xi'<\xi_1$,
\be
\fl
g(\xi,\xi')=I_<K_>-\frac{\mu_1\xi_1K_1^2+\mu_2\xi_2K_2^2-\mu_1\mu_2\xi_1\xi_2
K_1K_2(K_2I_1-K_1I_2)}{\Delta}II_\prime,\ee
where
\be
\Delta
=(1+\mu_1\xi_1K_1I_1)(1+\mu_2\xi_2K_2I_2)-\mu_1\mu_2\xi_1\xi_2
I_1^2K_2^2,
\ee
and where we have used the abbreviations
$I_a=I_\zeta(k\xi_a)$, $I=I_\zeta(k\xi)$, $I_\prime=I_\zeta(k\xi')$, etc.
For $\xi, \xi'>\xi_2$,
\be
\fl
g(\xi,\xi')=I_<K_>-\frac{\mu_1\xi_1I_1^2+\mu_2\xi_2I_2^2+\mu_1\mu_2\xi_1\xi_2
I_1I_2(I_2K_1-I_1K_2)}{\Delta}KK_\prime,\ee
and for $\xi_1$, $\xi_2$, for $\xi_1<\xi, \xi'<\xi_2$,
\bea
g(\xi,\xi')&=&I_<K_>
-\frac{\mu_2\xi_2K_2^2(1+\mu_1\xi_1K_1I_1)}\Delta II_\prime\nn\\
&&\mbox{}-\frac{\mu_1\xi_1I_1^2(1+\mu_2\xi_2K_2I_2)}\Delta KK_\prime
+\frac{\mu_1\mu_2\xi_1\xi_2I_1^2K_2^2}\Delta(IK_\prime+KI_\prime).
\eea

In the $\xi_0\to\infty$ limit, the UAE gives, for $\xi_1<\xi,\xi'<\xi_2$
($a=\xi_2-\xi_1$)
\bea
\fl\xi_0g(\xi,\xi')&\to&\frac1{2\kappa}\rme^{-\kappa|\xi-\xi'|}
+\frac1{2\kappa\tilde\Delta}\bigg[\frac{\mu_1\mu_2}{4\kappa^2}2\cosh\kappa
(\xi-\xi')\nn\\
\fl&&\quad\mbox{}-\frac{\mu_1}{2\kappa}\left(1+\frac{\mu_2}{2\kappa}\right)
\rme^{-\kappa(\xi+\xi'-2\xi_2)}
-\frac{\mu_2}{2\kappa}\left(1+\frac{\mu_1}{2\kappa}\right)
\rme^{\kappa(\xi+\xi'-2\xi_1)}\bigg],
\eea
with
\be
\tilde\Delta=\left(1+\frac{\mu_1}{2\kappa}\right)
\left(1+\frac{\mu_2}{2\kappa}\right)\rme^{2\kappa a}-\frac{\mu_1\mu_2}{4\kappa^2},
\ee
which is exactly the expected result.  The same holds in the other two regions.

\subsection{Force on two-plate system}
In general, we have two alternative forms for the force on the two-plate
system:
\be
\mathcal{F}=-(\partial_{\xi_1}+\partial_{\xi_2})\frac12\int\frac{\rmd\zeta\,
\rmd^2k}{(2\pi)^3}\ln\Delta,\ee
which is equivalent to
\be
\mathcal{F}=\int \rmd\xi\int\frac{\rmd\zeta\,\rmd^2k}{(2\pi)^3}
\hat\zeta^2g(\xi,\xi).
\ee
From either of these two methods, we find the
gravitational force on the Casimir energy to be in the $\xi\to\infty$ limit
\be
\mathcal{F}=-\frac1{4\pi^2}\int_0^\infty \rmd\kappa\,\kappa^2 \ln\Delta_0,
\quad
\Delta_0=\rme^{-2\kappa a}\tilde\Delta.
\ee
Explicitly,
\bea
\fl\mathcal{F}&=&\frac1{96\pi^2 a^3}\int_0^\infty \rmd 
y\,y^3\frac{1+\frac1{y+\mu_1a}
+\frac1{y+\mu_2a}}{\left(\frac{y}{\mu_1a}+1\right)
\left(\frac{y}{\mu_2a}+1\right)\rme^y-1}\nn\\
\fl&&\quad\mbox{}-
\frac1{96\pi^2 a^3}\int_0^\infty \rmd y\,y^2\left[\frac1{\frac{y}{\mu_1a}
+1}+\frac1{\frac{y}{\mu_2a}+1}\right]=
-(\mathcal{E}_c+\mathcal{E}_{d1}+\mathcal{E}_{d2})
,\eea
which is just the negative of the Casimir energy of the two semitransparent
plates.
The divergent terms are just the sum of the Casimir energies of each plate
separately, \eref{oneplate}, which serve to 
 simply renormalize the mass/area of each plate:
\be
E_{\rm total}=m_1+m_2+\mathcal{E}_{d1}+\mathcal{E}_{d2}+\mathcal{E}_c
=M_1+M_2+\mathcal{E}_c,
\ee
and thus the gravitational force on the entire apparatus obeys the
equivalence principle
\be
g\mathcal{F}=-g(M_1+M_2+\mathcal{E}_c).
\ee
Saharian et al.~\cite{Saharian:2003fd} earlier reached a similar conclusion, 
but only for the finite part of the energy.

\section{Conclusions}
\begin{itemize}

\item We have found, after a certain confusion, 
an extremely simple answer to how
Casimir energy gravitates: just like any other form of energy,
\be
\frac{F}A=-g\mathcal{E}_c.
\ee
This result is independent of the orientation of the Casimir apparatus
relative to the gravitational field.  This refutes the claim sometimes
attributed to Feynman that virtual photons do not gravitate.

\item Although gravitational energies have a certain ill-defined character,
being gauge- or coordinate-variant, this result is obtained for a Fermi
observer, the relativistic generalization of an inertial observer.
\item This conclusion is supported by an explicit calculation in Rindler
coordinates, describing a uniformly accelerated observer.  This demonstrates,
quite generally, that the total Casimir energy, including the divergent
parts, which renormalize the masses of the plates,
possesses the gravitational mass demanded by the equivalence principle.

\item  
New developments of this work are in progress, and will be described in part
in the contribution to this proceedings by K V  Shajesh.
\end{itemize}
\ack
This work was supported in part by Collaborative Research Grants from
the US National Science Foundation (PHY-0554849 and PHY-0554926)
and in part by the
US Department of Energy (DE-FG02-04ER41305).  We are grateful for Michael
Bordag arranging such a fruitful QFEXT07 workshop.


\section*{References}
\begin{thebibliography}{[99]}

\bibitem{Jaffe:2005vp}
  Jaffe R L 2005
{\it  Phys.\ Rev.\  D} {\bf 72} 021301
  [arXiv:hep-th/0503158]

\bibitem{Boyer:1968uf}
  Boyer T H 1968
{\it  Phys.\ Rev.} {\bf 174} 1764 
  
\bibitem{Graham:2003ib}
 Graham N, Jaffe R L, Khemani V, Quandt M, Schroeder O and Weigel H 2004
 {\it Nucl.\ Phys.\ B} {\bf 677} 379
  [arXiv:hep-th/0309130], and references therein.
 
\bibitem{barton} Barton G 2004 {\it J. Phys. A} {\bf 37} 1011; 
2001 {\it J. Phys. A}
{\bf 34} 4083

\bibitem{Deutsch:1978sc}
Deutsch D and Candelas P 1979
  %
{\it  Phys.\ Rev.\ D} {\bf 20} 3063

\bibitem{Candelas:1981qw}
 Candelas P 1982
  %
{\it  Ann.\ Phys.\ (N.Y.)} {\bf 143} 241;
1986  {\it  Ann.\ Phys.\ (N.Y.)} {\bf 167} 257

\bibitem{CaveroPelaez:2006kq}
  Cavero-Pel\'aez I, Milton K A and Wagner J 2006
{\it  Phys.\ Rev.\  D} {\bf 73} 085004 
[arXiv:hep-th/0508001]
  

\bibitem{CaveroPelaez:2006rt}
Cavero-Pel\'aez I, Milton K A and Kirsten K 2007
{\it  J.\ Phys.\ A}  {\bf 40} 3607
  [arXiv:hep-th/0607154]

\bibitem{brown} Brown L S and Maclay G J 1969 
{\it Phys.\ Rev.} {\bf 184} 1272
  
\bibitem{casbook} Milton K A 2001 
{\em The Casimir Effect: Physical Manifestations
of Zero-Point Energy} (World Scientific, Singapore), Sec.~11.1

\bibitem{Fulling:2007xa}
Fulling S A, Milton K A, Parashar P, Romeo A, Shajesh K V and Wagner J 2007
{\it  Phys.\ Rev.\  D} {\bf 76}  025004
  [arXiv:hep-th/0702091]

\bibitem{js} Schwinger J 1970
{\it Particles, Sources, and Fields} (Addison-Wesley, Reading, MA)

\bibitem{calloni} Calloni E, Di Fiore L, Esposito G, Milano L and Rosa L 2002
{\it  Phys.\ Lett.\  A} {\bf 297} 328
  [arXiv:quant-ph/0109091]

\bibitem{bimonte} Bimonte G, Calloni E, Esposito G, and Rosa L 2006
{\it Phys.\ Rev.\ D} {\bf 74} 085011 [arXiv:hep-th/0606042]

\bibitem{jaekel} Jaekel M T and Reynaud S 1993
{\it Journal de Physique I} {\bf3} 1093

\bibitem{weinberg} Weinberg S 1972 {\it Gravitation and Cosmology: Principles 
and Applications of the General Theory of Relativity} (Wiley, New York)



\bibitem{marzlin}
Marzlin K-P 1994 {\it Phys.\ Rev.\ D} {\bf 50} 888

\bibitem{Bimonte:2007zt}
 Bimonte G, Calloni E, Esposito G and Rosa L 2007
{\it  Phys.\ Rev.\  D} {\bf 76}  025008
  [arXiv:hep-th/0703062]

\bibitem{Milton:2007ar}
  Milton K A, Parashar P, Shajesh K V and Wagner J 2007
{\it  J.\ Phys.\ A}  {\bf 40}  10935
  [arXiv:0705.2611 [hep-th]]




\bibitem{Saharian:2003fd}
Saharian A A, Davtyan R S and Yeranyan A H 2004
{\it  Phys.\ Rev.\  D} {\bf 69}  085002
  [arXiv:hep-th/0307163]


\end{thebibliography}
\end{document}